\documentclass[aps,prl,twocolumn,showpacs,superscriptaddress]{revtex4-1}  
\usepackage{graphicx}  
\usepackage{dcolumn}   
\usepackage{bm}        
\usepackage{amsmath}

\begin{document}

\title{Quantum state tomography by continuous measurement and compressed sensing}

\affiliation{Center for Quantum Information and Control, College of Optical Sciences and Department of Physics, University of Arizona, Tucson, Arizona 85721, USA}
\affiliation{Center for Quantum Information and Control, Department of Physics and Astronomy, University of New Mexico, Albuquerque, New Mexico 87131, USA}

\author{A.~Smith}\affiliation{Center for Quantum Information and Control, College of Optical Sciences and Department of Physics, University of Arizona, Tucson, Arizona 85721, USA}
\author{C. A.~Riofr\'io}\affiliation{Center for Quantum Information and Control, Department of Physics and Astronomy, University of New Mexico, Albuquerque, New Mexico 87131, USA}
\author{B. E.~Anderson} \affiliation{Center for Quantum Information and Control, College of Optical Sciences and Department of Physics, University of Arizona, Tucson, Arizona 85721, USA}
\author{H.~Sosa-Martinez}\affiliation{Center for Quantum Information and Control, College of Optical Sciences and Department of Physics, University of Arizona, Tucson, Arizona 85721, USA}
\author{I. H.~Deutsch}\affiliation{Center for Quantum Information and Control, Department of Physics and Astronomy, University of New Mexico, Albuquerque, New Mexico 87131, USA}
\author{P.S.~Jessen}\affiliation{Center for Quantum Information and Control, College of Optical Sciences and Department of Physics, University of Arizona, Tucson, Arizona 85721, USA}

\vskip 0.25cm

\date{\today}

\begin{abstract}
The need to perform quantum state tomography on ever larger systems has spurred a search for methods that yield good estimates from incomplete data. We study the performance of compressed sensing (CS) and least squares (LS) estimators in a fast protocol based on continuous measurement on an ensemble of cesium atomic spins. Both efficiently reconstruct nearly pure states in the 16-dimensional ground manifold, reaching average fidelities $\bar{\mathcal F}_{CS} = 0.92$ and $\bar{\mathcal F}_{LS} =0.88$ using similar amounts of incomplete data. Surprisingly, the main advantage of CS in our protocol is an increased robustness to experimental imperfections.
\end{abstract}

\maketitle

Recovering a full description of a complex system from limited information is a central problem in science and engineering. In physics one often seeks to estimate an unknown quantum state based on measurement data \cite{Paris2004}, generally a formidable challenge for large systems given that $\mathcal{O}(d^2)$ real parameters are needed to describe arbitrary states in a $d$-dimensional Hilbert space. In quantum information science, however, the states of interest are nearly pure and can be described by $\mathcal{O}(d)$ parameters.  Algorithms that make use of this prior information to obtain good estimates from a reduced number of measurements fall under the general heading of compressed sensing \cite{Candes2008}, a family of techniques used in signal processing tasks that range from movie recommendation to earthquake analysis. Gross et al.  \cite{Gross2010,Flammia2012} have developed one such algorithm that gives good estimates of nearly pure quantum states in a $d$-dimensional Hilbert space from the expectation values of $\mathcal{O}(d$\hspace{0.5mm}log\hspace{0.5mm}$d)$ orthogonal observables, a substantial saving when $d$ is large. This algorithm was recently benchmarked against a standard maximum likelihood estimator in an experiment with photonic qubits  and the two were found to yield similar results \cite{Liu2012}. Generalization to process tomography has led to similar improvements when the process is close to unitary \cite{Shabani2011a}.

\begin{figure}
\begin{center}
\includegraphics[scale=0.1]{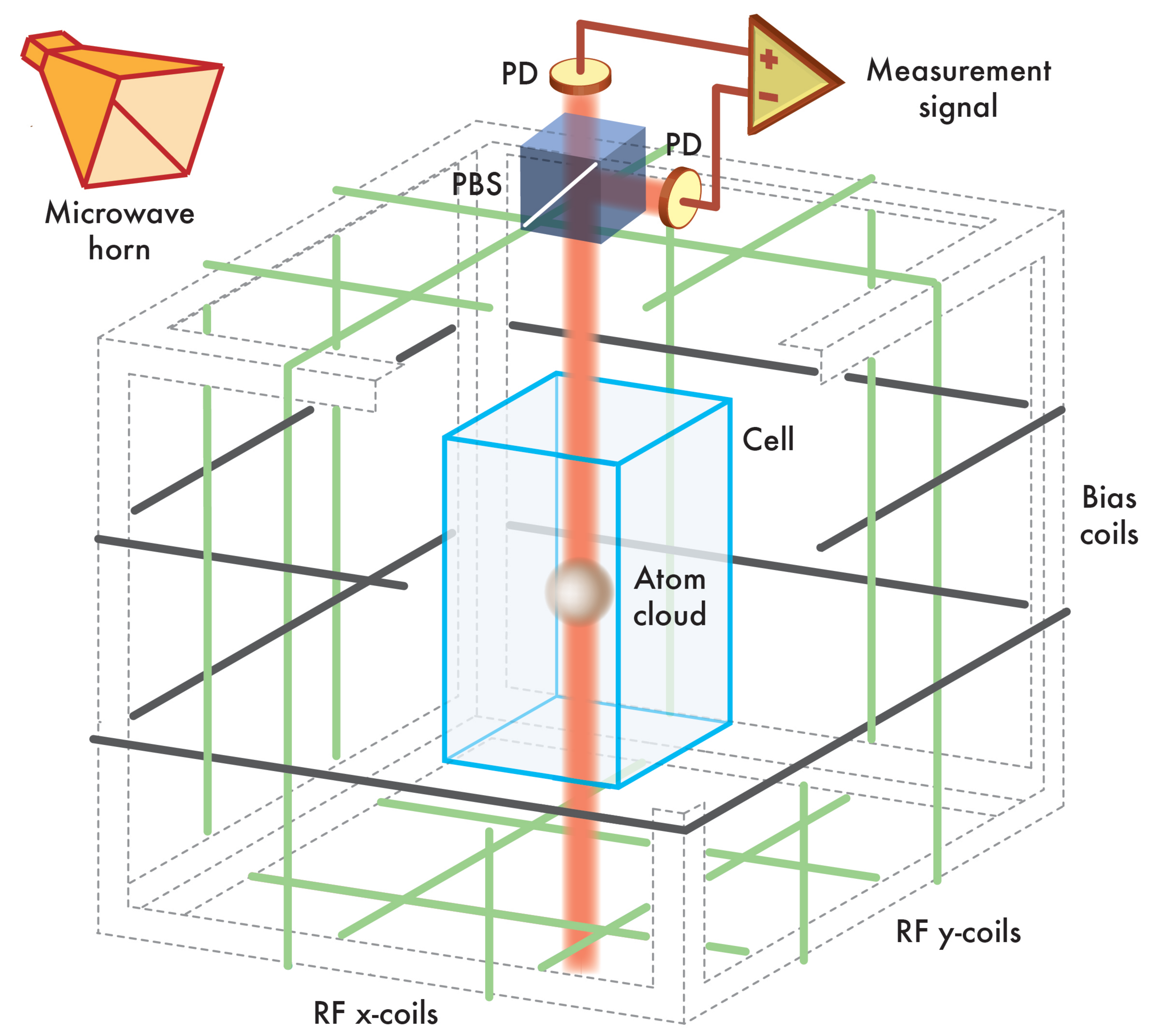}
\caption[Quantum Tomography by Continuous Measurement]{\label{fig:TomoSchematic} (Color online) Schematic of the experiment.  An ensemble of identically prepared cesium atoms is probed with an optical beam and polarimeter to obtain a continuous measurement of the spin observable $f_z$ in the $f=3$ hyperfine state.  The atoms sit at the center of a plexiglass cube that supports coil pairs used to apply bias and rf magnetic fields.  The upper corner is cut-away in the illustration to show the atom cloud and vacuum cell. A horn antenna radiates the microwave field that is also required for full dynamical control.}
\end{center}
\end{figure}

In this work we study the laboratory performance of quantum state reconstruction based on compressed sensing (CS) and least-squares \cite{James2001}  (LS) estimators in the context of continuous measurement. Our physical testbed consists of the 16-dimensional hyperfine manifold of magnetic sublevels in the electronic ground state of atomic cesium. The data required for quantum tomography is gathered by performing a weak (nonprojective) continuous measurement on an ensemble of atoms while dynamically evolving their state with known driving fields \cite{ Riofrio2011,Deutsch2010, Silberfarb2005, Smith2006}. This approach differs substantially from conventional quantum tomography in that the measurement record contains information about the expectation values of a {\it continuum} of nonorthogonal observables instead of a discrete orthogonal set. We find that the CS and LS estimators both achieve high fidelity reconstruction of nearly pure states from similar amounts of incomplete and noisy data, but CS appears to be significantly more robust against imperfections in the experimental implementation. From a practical perspective our approach offers the advantage of very fast data collection, as the combination of continuous joint measurement and statistical averaging over many atoms allows us to obtain an informationally complete measurement record from a single ensemble containing many copies of the quantum state.  This makes it feasible to measure and compare the fidelity of the CS and LS algorithms across a large sample of test states.

We begin our experiment (Fig. 1) with an ensemble of $\sim$$10^6$ cesium atoms captured and cooled in a magneto-optic trap and optical molasses, released  into free fall, and optically pumped into a pure hyperfine state $|f = 4, m_{f} = 4\rangle$. Dynamical evolution of the single-atom quantum state is driven with a combination of four magnetic fields: a static bias field producing a Zeeman splitting of 1 MHz, two orthogonal rf fields oscillating at 1 MHz that drives Larmor precession independently in the $f = 3$ and $f = 4$ hyperfine manifolds, and a $\mu$w field resonant with the $|f=3,m_{f}=3\rangle \leftrightarrow |f=4,m_{f}=4\rangle$ transition at $\sim9.2$ GHz \cite{Merkel2008}. Background magnetic fields are suppressed to less than $\sim60$ $\mu$G, corresponding to an uncertainty in the Zeeman splitting of  $\sim20$ Hz \cite{Smith2011}. Arbitrary pure test states can be prepared with fidelities $\mathcal{F} \geq 0.99$, by evolving the initial state using rf and $\mu$w fields of fixed amplitude and frequency, and with computer optimized modulation of the phases \cite{Smith2012}. During tomography the atoms are similarly driven with rf and $\mu$w fields of fixed amplitude and frequency, and with phases that are modulated in piecewise steps of $15  \mu$s and $10  \mu$s respectively.  These phases are chosen at random to generate a set of three phase modulation waveforms; this one set is subsequently used in every trial of the experiment.

Our continuous measurement is performed with a probe beam tuned $730$ MHz below the $6{\rm S}_{1/2} (f = 3) \leftrightarrow 6{\rm P}_{1/2} (f = 4)$ transition of the D1 line where the light shift has minimal effect on the dynamics in the hyperfine ground manifold.  Rotation of the probe polarization provides a measurement of the collective spin projection of the atoms, $F_{z}=\sum_{i}{f^{(i)}_{z}}$, where $z$ is the direction of probe propagation and $f_z$ is a single-atom operator associated with the angular momentum in the $f=3$ hyperfine state.  The atom-probe coupling is sufficiently weak that the entangling effect of measurement backaction is negligible, and the ensemble is well approximated by a product state at all times.  In this situation the measurement record is of the form $M(t)=K\langle f_{z}(t)\rangle+\sigma W(t)$, were $K$ is proportional to the optical depth of the ensemble, and $W(t)$ is Gaussian white noise with variance $\sigma^{2}$ representing probe shot noise \cite{Deutsch2010}.  In practice, the contribution from probe shot noise is negligible compared to systematic errors in the expectation value $ \langle f_{z}(t)\rangle$ caused by imperfections in the dynamics.

In a given trial of the experiment, our objective is to find an estimate ${\bar \rho}$ of the initial state $\rho_{0}$ based on the experimentally observed $M(t)$ and the known dynamics. Working in the Heisenberg picture, we do this by discretizing the measurement record and associated time-dependent observable into time series, $M_{i}=M(t_{i})$  and $O_{i}=O(t_{i})$, where $O_{0}=f_{z}$. We then estimate ${\bar \rho}$ using two algorithms:
\begin{gather}
\makebox[-80pt][r]{\underline{Least Squares (LS)}} \\
{\bar \rho} = \mbox{arg min}_{\rho}\sum_{i}[M_{i}-KTr(\rho O_{i})]^{2} \nonumber \\
\mbox{subject to } Tr(\rho)=1 \mbox{ , } \rho^{\dagger}=\rho \mbox{ , } \rho \geq 0 \nonumber
\end{gather}
\begin{gather}
\makebox[-20pt][r]{\underline{Compressed Sensing (CS)}} \\
{\bar \rho} = \mbox{arg min}_{\rho}Tr(\rho) \nonumber \\
\mbox{subject to }\sum_{i}[M_{i}-KTr(\rho O_{i})]^{2} \leq \epsilon \mbox{ , } \rho^{\dagger}=\rho \mbox{ , } \rho \geq 0 \nonumber \\
\mbox{Renormalize so } Tr({\bar \rho})=1 \nonumber
\end{gather}
In the CS algorithm, minimizing $Tr(\rho)$ is a known heuristic for minimizing the rank (maximizing the purity) of $\rho$ \cite{Candes2011,Gross2011}. The constant $\epsilon$ depends on measurement uncertainty and is chosen empirically as discussed below. In both cases ${\bar \rho}$ must be normalized and physical, i.e., Hermitian with positive eigenvalues.

To solve these optimization problems, the unknown state of the system is parametrized by the set of $d^{2}$ real numbers $\{r_{\alpha}\}$, such that $\rho=\sum_{\alpha=0}^{d^{2}-1}r_{\alpha}E_{\alpha}$, where $\{E_{\alpha}\}$ is an orthonormal basis of Hermitian, traceless operators, and $E_{0} = I/\sqrt{d}$. With this parametrization the distance squared between observed and predicted measurements can be expressed in terms of the unknown parameters $\{r_{\alpha}\}$ as $\Delta=\sum_{i}[M_{i}-K\sum_{\alpha=0}^{d^{2}-1}r_{\alpha}Tr(O_{i}E_{\alpha})]^{2}$. In CS we minimize $Tr(\rho)=r_{0}\sqrt{d}$ subject to $\Delta \leq \epsilon$, while in LS we minimize $\Delta$ directly. Both algorithms present standard convex problems~\cite{Boyd2004}, which we solve in MATLAB using a freely available package~\cite{Grant2011}.      

\begin{figure*}
\begin{center}
\includegraphics[scale=0.54]{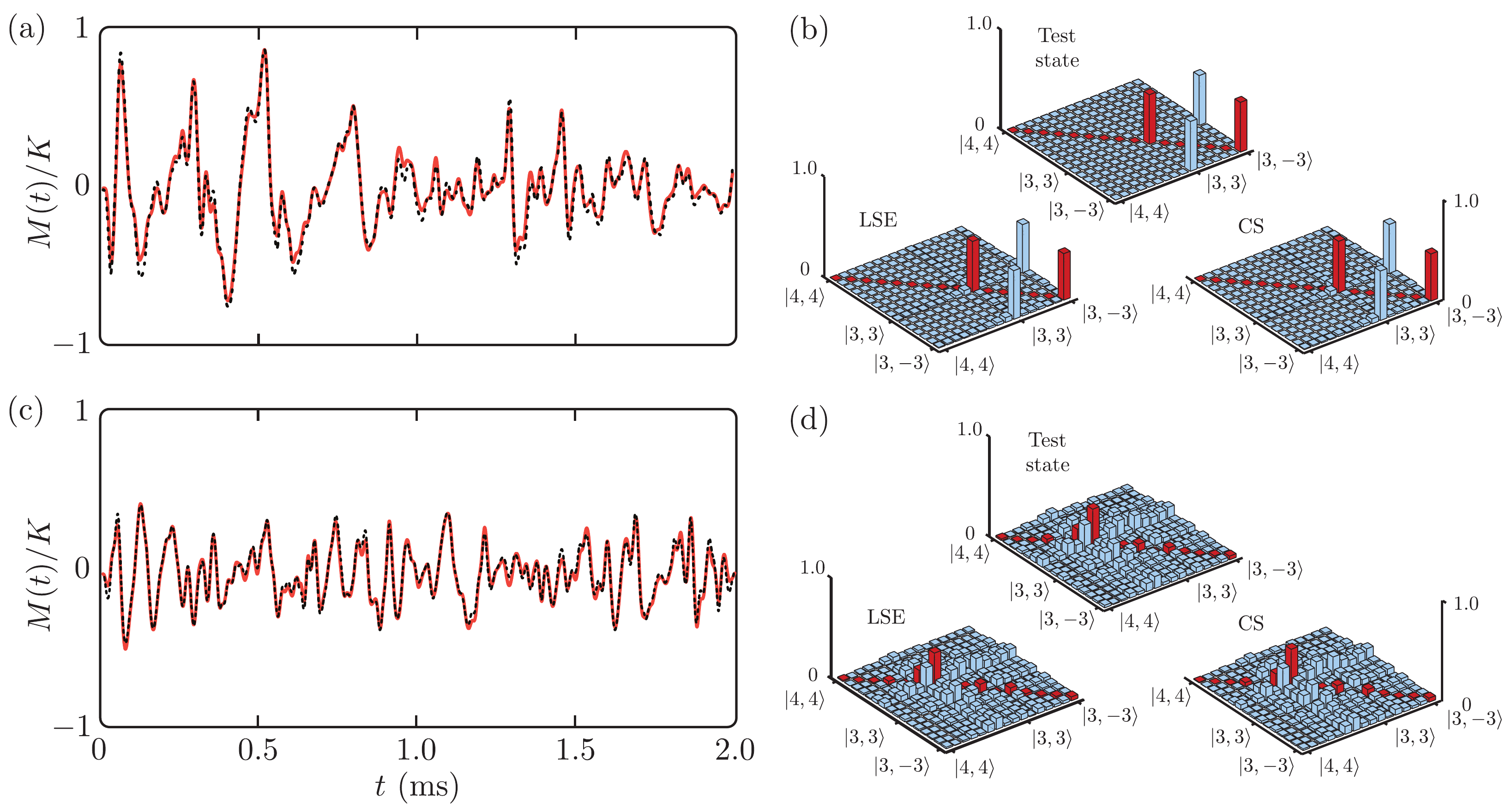}
\caption[Measurement Records and Estimated States]{\label{fig:ReconExamples} (Color online) (a) Observed (black dotted line) and predicted (red solid line) measurement records for the test state $|\psi_{0}\rangle=(|3,3\rangle+|3,-3\rangle)/\sqrt{2}$. (b) Density matrices for the test state in (a), and for the reconstructed states obtained via CS and LS.  (c) Observed (black dotted line) and predicted (red solid line) measurement records for a random test state. (d) Density matrices for the test state in (c), and for the reconstructed states obtained via CS and LS.  All density maurice are shown in the basis of hyperfine magnetic sublevels, $|f,m_{f}\rangle$, arranged in order $\{|3,-3\rangle...|3,3\rangle,|4,-4\rangle...|4,4\rangle\}$.  Red (dark) color indicates populations and blue (light) color indicates coherences; for the coherences only the absolute values are shown}
\end{center}
\end{figure*}

From an experimental perspective, the main challenge is to generate a set of observables $\{O_{i}\}$ that are known with sufficient accuracy. Under ideal conditions (no decoherence or experimental imperfections), theoretical simulation shows that sets $\{O_{i}\}$ generated by different random dynamics approach informational completeness in roughly equal time. For our parameters (rf and $\mu$w amplitude and frequency) and phase modulation waveforms, this takes a few ms. Once a set of parameters and waveforms have been chosen, careful modeling of the experiment is required to determine the observables that are actually measured in a particular run. This involves independent determination of a number of experimental variables, and the numerical integration of a master equation that accounts for decoherence and inhomogeneous driving fields (see \cite{Riofrio2011} for details on modeling and simulation). Figure 2 shows the observed and predicted measurement records for two test states, and illustrates the basic traits required for state estimation: the observed and predicted measurement records are in excellent agreement, and the measurement records from different states are distinct.  Fundamentally, this is what allows the measurement records to serve as identifiable quantum ``fingerprints".

For a more comprehensive evaluation of the LS and CS estimators, we look at their performance averaged over a set of 49 test states chosen randomly according to the Haar measure~\cite{Mezzadri2007}.  One of the test states is selected at random and used to establish a threshold for CS by chosing the $\epsilon$ that maximizes the fidelity of the state estimate for a measurement record of given length.  We find empirically that the best choice of $\epsilon$ increases linearly with the length of the measurement record, which is expected because the distance measure $\Delta$ grows in proportion to the number of measurements included in the summation.  This calibration state is subsequently discarded. The measurement records for the remaining test states are analyzed using both CS and LS, and the resulting estimates ${\bar \rho}$ are compared to the known inputs $\rho_{0}=|\psi_{0}\rangle \langle \psi_{0}|$.  We have found empirically that different choices of calibration state do not lead to significant changes in $\epsilon$, or in the corresponding fidelities when the remaining states are estimated. Figure \ref{fig:ReconExamples}b,d shows input and reconstructed states for the measurement records in Fig. \ref{fig:ReconExamples}a,c with fidelities ${\mathcal F}_{CS} = \langle \psi_{0}|{\bar \rho}|\psi_{0}\rangle = 0.98(3)$ and $\mathcal{F}_{LS} = 0.97(3)$ for the first example, and $\mathcal{F}_{CS} = 0.96(3)$ and $\mathcal{F}_{LS} = 0.90(3)$ for the second. Here and elsewhere, numbers in parenthesis indicate the uncertainty (one standard deviation) on the last digit. It is worth noting that we generally achieve higher fidelity for states that have support on only a few magnetic sublevels (e. g., the example in Fig. \ref{fig:ReconExamples}b), though the reasons for this are not yet clear. Averaged over a sample of 48 Haar random test states, we find $\bar{\mathcal F}_{CS} = 0.917(4)$ and $\bar{\mathcal F}_{LS} =0.880(4)$. The corresponding infidelities indicate that CS outperforms LS by ~30\% for this data set.

\begin{figure*}
\begin{center}
\includegraphics[scale=0.67]{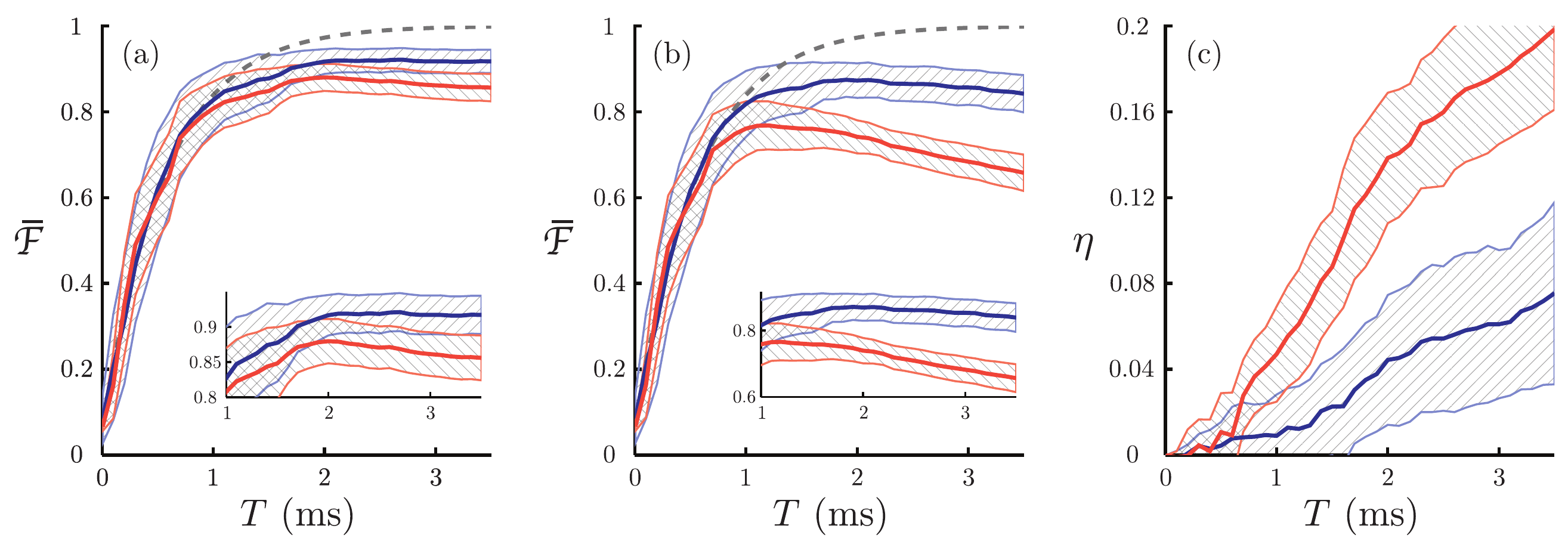}
\caption[Performance of CS and LS algorithms]{\label{fig:Performance} (Color online) (a) Fidelity based on measurements at times $t\leq T$ only, as a function of $T$.  Solid lines are the average fidelity for a sample of 48 random test states reconstructed via CS (dark blue) or LS (light red), the dashed line is a fit to the data for times $T < 1$ ms.  Cross-hatched areas are $\pm$ one standard deviation for the fidelities of the test states; this is an indication of the deviation from the average that typically occur in a single state reconstruction.  (b) Same as (a), except that a deliberate error was introduced in the theoretical model of the experiment.  (c) Error penalty $\eta(T)={\bar {\mathcal F}}^{a}(T)-{\bar {\mathcal F}}^{b}(T)$ incurred due to this error.}
\end{center}
\end{figure*}

The performance of a state estimation algorithm is measured not only by fidelity but also in terms of efficiency, i.e., how much data is necessary to reconstruct the quantum state. In conventional tomography the efficiency is quantified by the number of orthogonal observables and the number of measurements required per observable. In contrast, our protocol measures a slightly different observable every microsecond. In this situation the set $\{O_{i}\}$ spans the space of observables after a very short time, but the degree of certainty with which we have measured the corresponding expectation values increases much more slowly. A reasonable way to quantify efficiency is, therefore, to reconstruct the state based on measurements at times $t \leq T$, and observe how the fidelity improves as $T$ increases. Figure \ref{fig:Performance}a shows our average fidelities  $\bar{\mathcal F}_{CS}(T)$ and $\bar{\mathcal F}_{LS}(T)$, together with fits to an exponential rise, $\bar{\mathcal F}(T)=(15/16)(1-{\rm exp}[-T/\tau])+1/16$, for $T < 1$ ms. This data contains information about several aspects of our protocol. First, for our parameters the peak fidelities (quoted in the previous paragraph) are reached at $T\sim2$ ms. Second, the CS and LS fidelities rise with similar time constants, $\tau_{CS} = 0.56(1)$ ms and $\tau_{LS} = 0.57(2)$ ms. To put these values in perspective, it is instructive to compare to LS of a maximally mixed state. Applying our protocol to simulated measurements \cite{Riofrio2011}, we see exponential rise in the average fidelity for pure as well as maximally mixed states, but with a time constant that is approximately four times larger in the latter case. This indicates that both the CS and LS algorithms make good pure-state estimates from measurement records that contain much less information than required to describe arbitrary mixed states. In the case of LS we believe this happens due to the positivity constraint $\rho_{0} \geq 0$, which sharply limits the states consistent with the measurement record in the vicinity of a pure state \cite{Byrd2003,Kimura2003} - exactly the regime where CS is also efficient.

The data in Fig. \ref{fig:Performance} also contains information about a third aspect of CS and LS: robustness in the presence of noise and imperfections. It is well known that experimental imperfections can cause systematic errors in the measurement settings and lead to poor quality tomography as a result. Recent work has explored  methods to reject those data sets that are contaminated by errors beyond quantum statistical uncertainty, using concepts analogous to an ``entanglement witness''~\cite{Moroder2012} or other statistical fitness tests~\cite{Langford2012}.  In our protocol, systematic errors in the measurement settings appear when the actual versus modeled dynamics lead to incorrect assumptions about the measured $\{O_{i}\}$. Such incorrect modeling of the true experimental conditions will results in fidelity that {\em decreases} as a function of time as dynamical errors are cumulative.  LS is particularly vulnerable to systematic errors in the measurement settings  $\{O_{i}\}$ because it minimizes the distance between observed and predicted measurements, even when one or both cannot be trusted beyond a certain level. By comparison, CS only requires the distance to fall below some threshold $\epsilon$, and can produce more reliable results by ignoring meaningless variations in either. This is supported by Fig. \ref{fig:Performance}a, where the CS and LS fidelities rise with similar time constants at short times when the experiment is well modeled, peak roughly simultaneously, and then plateau (CS) or decline (LS) at later times. 

The superior robustness of CS is even more pronounced in poor quality data sets for which we achieve generally poor state estimation, which typically occurs due to accidental miscalibration of one or more experimental variables.  In practice it is difficult to generate poor quality data sets in a controlled manner, and it is better to start from a high quality data set and deliberately introduce a known error in the model. Figure \ref{fig:Performance}b shows an example in which we have left out an average that accounts for spatial inhomogeneity of a parameter in the experiment. Figure \ref{fig:Performance}c shows the resulting error penalty (increase in infidelity) when going from Fig. \ref{fig:Performance}a to Fig. \ref{fig:Performance}b, $\eta(T)=\bar{\mathcal F}^{a}(T)-\bar{\mathcal F}^{b}(T)$. In this example the error penalty for CS is roughly a factor of 3 less than for LS at $T \sim 2$ ms when peak fidelities are reached. Inaccurate calibration of an inhomogeneity is one of the more likely imperfections to occur in our experiment, but the trend seen here occurs also when we introduce other types of imperfections.

Our implementation of CS and LS in continuous-time quantum tomography raises new questions about the relative performance of these algorithms in a realistic experimental setting.  We have seen that in our protocol, the {\em rate} at which the fidelity of state estimation increases is essentially the same for CS and LS,  whereas the CS estimator is more robust to experimental imperfections.  The role of positivity, the nature of experimental noise and imperfections, and the overcompleteness of the measurement settings~\cite{Burgh2008} may all play important roles in explaining these features and will be explored in future research.  Our continuous-time protocol also opens up new directions for tomography in situations where the state of the system is well known but the dynamics is not. In that case the measurement record contains information about the process, and could potentially be used as the basis for new protocols aimed at fast Hamiltonian tomography or other forms of parameter estimation \cite{Schirmer2009, Franco2009, Shabani2011b,  Gammelmark2012}.

This work was supported by the US National Science Foundation Grants PHY -0903692, 0903930, 0903953.

\end{document}